# A spin triplet supercurrent through the half-metallic ferromagnet $CrO_2$

R. S. Keizer[1], S. T. B. Goennenwein[1]†, T. M. Klapwijk[1], G. Miao[2,3], G. Xiao[3] & A. Gupta[2]

[1]*Kavli Institute of NanoScience, Faculty of Applied Sciences, Delft University of Technology, 2628 CJ, Delft, The Netherlands.* [2]*MINT Center, University of Alabama, Tuscaloosa, Alabama 35487, USA.* [3]*Physics Department, Brown University, Providence, Rhode Island 02912, USA. †Present address: Walther-Meißner-Institut, Bayerische Akademie der Wissenschaften, Walther-Meißner-Straße 8, D-85748, Garching, Germany.*

**In general, conventional superconductivity should not occur in a ferromagnet, though it has been seen in iron under pressure[1]. Moreover, theory predicts that the current is always carried by pairs of electrons in a spin singlet state[2], so conventional superconductivity decays very rapidly when in contact with a ferromagnet, which normally prohibits the existence of singlet pairs. It has been predicted that this rapid spatial decay would not occur if spin triplet superconductivity could be induced in the ferromagnet[3,4]. Here we report a Josephson supercurrent through the strong ferromagnet $CrO_2$, from which we infer that it is a spin triplet supercurrent. Our experimental set-up is different from those envisaged in the earlier predictions, but we conclude that the underlying physical explanation for our result is a conversion from spin singlet pairs to spin triplets at the interface. The supercurrent can be switched with the direction of the magnetization, analogous to spin valve transistors, and therefore could enable magnetization-controlled Josephson junctions.**

In our experiment we realized a sample (Fig. 1b) in which two *s*-wave superconductors, made out of NbTiN (ref. 5), are coupled by the conducting ferromagnet $CrO_2$, a material well known from magnetic recording tapes[6]. Using electron-beam lithography, sputtering and lift-off, two 'T'-shaped NbTiN electrodes with a relatively large mutual distance of 0.3–1 μm are patterned on top of the $CrO_2$ (Fig. 1c). On cooling the sample to temperatures between 1 and 10 K, we find that the current between the two electrodes, which can only pass through the ferromagnetic $CrO_2$ film, is a supercurrent (Fig. 2a). We also find that with increasing temperatures the maximum supercurrent, $I_c$, decreases, and disappears at a temperature comparable to the superconducting transition temperature, $T_c$, of the thin NbTiN film (Fig. 2b). The observation of a supercurrent through a ferromagnet has been reported before[7,8], but only for very weak ferromagnets and over significantly shorter distances.

The ferromagnet we use, $CrO_2$, is a so-called half-metallic ferromagnet[6]. The electronic transport is metallic for the spin-up electrons, while it is insulating for the spin-down electrons, a property which is entirely due to the band structure of the material, schematically shown in Fig. 1a. Some details of the electronic structure are still under debate, but $CrO_2$ is assumed to be a self doped double exchange ferromagnet with a gap (of ~2 eV) in the spin-down density-of-states at the Fermi level[9]. The material has a Curie temperature $T_{Curie} \approx 390$ K and is metallic at low temperatures[10], with a resistivity of about 8.9 μΩ cm at 1.6 K. It has been experimentally demonstrated that, as expected, the spin polarization is close to 100% (refs 11, 12). The saturation magnetization is equal to two Bohr magnetons ($2\mu_B$) per unit cell. Important for our experiments is that the magnetic behaviour (switching/rotation of the magnetization direction) has been shown to be single-domain-like, even for macroscopic films[13].

From magnetotransport measurements at temperature $T=4.2$ K (in which the resistance of the film is measured as a function of an external, in-plane magnetic field), we find the $CrO_2$ films to have a biaxial (cubic) magneto-crystalline symmetry in the plane of the film. This biaxial character appears as two switches in the resistance as a function of field, which enabled us to identify the directions of the easy axes of the ferromagnet: 30°, 150°, 210° and 330° to the crystallographic *c* axis. As shown in Fig. 1d, the junction is aligned along this axis, while the *a* axis and *b* axis are out-of-plane and in-plane, respectively.

In conventional superconductors—such as NbTiN used here—electrons are paired in so-called singlet Cooper pairs. In singlet pairs, an electron with spin up is paired with another electron with spin down. When two superconductors are coupled through a normal metal, a supercurrent will flow when the thickness of the normal layer is less than, or of the order of, the normal metal coherence length $\xi_N = \sqrt{\hbar D / k_B T}$, with Planck's constant $\hbar$, the diffusion constant for elastic scattering $D$, and Boltzmann's constant $k_B$. In normal metals, $\xi_N \approx 100$ nm is a measure for the length over which a Cooper pair looses its coherence, and is insensitive to the spin of the electrons forming the Cooper pair. If the normal metal is replaced by a conducting ferromagnet, as in our experiment, the electrons sense the magnetization and are pulled apart in energy depending on their spin orientation. $\xi_N$ must then be replaced by $\xi_F \approx \sqrt{\hbar D / k_B T_{Curie}}$. As typically $T \ll T_{Curie}$, $\xi_F$ in a ferromagnet is short, of the order of 1 nm. The reason is that a singlet Cooper pair (just as a normal metal) is symmetric for the spin directions. In contrast, a ferromagnet likes to have all spins pointing in the same direction. Several recent experiments have confirmed this short-length-scale picture[7,8].

However, Bergeret, Volkov and Efetov[3] have demonstrated theoretically that coherent triplet Cooper pairs—that is, pairs of electrons with both spins in parallel—can be induced in a ferromagnet in close proximity to a conventional singlet pair superconductor, given suitable conditions[14]. They also predict that the coherence length will be equal to $\xi_N$ of a normal metal, hence the name 'long range proximity effect'. These authors furthermore show[4] that triplet correlations are, unlike the anomalous order parameters in $^3$He (ref. 15) and unconventional superconductors[16,17], robust against impurity scattering. There are some indications for the existence of these long-range correlations[18–20], but an unequivocal experiment ruling out other explanations, such as the observation of a Josephson supercurrent through a fully spin-polarized ferromagnet, has not been reported. In our experiment, we observe such a supercurrent which prevails over very long length scales ~1 μm, that is much longer than expected for singlet correlations, and which characteristically depends on the orientation of the magnetization in the ferromagnet. Therefore, we attribute this long-range supercurrent to superconducting triplet correlations.

In conventional Josephson junctions, the supercurrent $I_s$ is given by $I_c \sin(\varphi_1-\varphi_2)$, meaning that the maximum value, the critical current $I_c$, is obtained for a phase difference $\varphi_1-\varphi_2=\pi/2$, with $\varphi_1$ the quantum phase of superconductor 1 and $\varphi_2$ the quantum phase of superconductor 2. The application of an external magnetic field, **H**, creates a position dependence of the phases $\varphi_1$ and $\varphi_2$, leading to a characteristic periodic dependence on magnetic induction, **B**, known as a Fraunhofer pattern[2] (in analogy to optical diffraction). We have applied a magnetic field in the plane of the $CrO_2$ film. Besides the conventional Fraunhofer pattern, we observe additional effects due to the finite magnetization, **M**, of the sample, as shown in Fig. 2c. As evident from the raw data shown in the inset, there are clear signs of hysteresis in the critical current as a function of the magnetic field. Owing to the biaxial symmetry of the magnetic system, the magnetization vector follows a different trajectory along the easy directions of the film (clockwise versus anticlockwise) in the up and down sweep of the magnetic field (Fig. 1d). This results in the magnetization in the upsweep lagging behind with respect to the magnetization in the downsweep, leading to an overall shift in the two sweeps (of the order of 45 mT). After removal of the hysteresis, clear oscillations are visible in the critical current, corresponding to the addition of one flux quantum through the junction area per period of the oscillation of 80 mT. This corresponds to an effective junction length of 240 nm (which is somewhat shorter than the actual length of the junction, 310 nm). Although the periodic dependence is quite analogous to the Fraunhofer pattern, there are clear deviations; that is, there is a minimum at the centre rather than a peak. We attribute this effect to the finite sample magnetization, which adds an offset to the flux density, and thus shifts the maximum of the pattern away from **H**=0.

The biaxial symmetry is also clearly evident from the following experiment. Figure 3 compares measurements obtained by first making a full magnetic field sweep (rotating the magnetization by 360°, 'major loop') and then a sweep that returns halfway ('minor loop'). These data are similar to those obtained in a giant magnetoresistance experiment with not the resistance but the supercurrent being changed by the magnetization. The relative change in $I_c$ of the magneto-switchable Josephson junction is in our case around 30%, and can be further increased by alignment of the junction along an easy axis. Hence, we find that the supercurrent is strongly linked to the magnetization **M** of the $CrO_2$ for a field **H** applied in the plane of the film (Fig. 1d). This double dependence on both **H** and **M** is a unique aspect of these ferromagnetically coupled Josephson junctions.

At present it is difficult to make a quantitative comparison of our measurements to theory. Comparing the temperature dependence of the critical current with the expressions following from the diffusive theory[21] leads to Thouless energies in the range from 30 to 80 μeV. There is one theoretical paper[22] in which triplet correlations are analysed for a geometry resembling our experiment, but only in the limit of one-dimensional ballistic motion, which is not applicable to our samples. Apart from this, a crucial ingredient of theoretical descriptions is the process and the strength of the singlet–triplet conversion. This process is generally believed to be caused by an interplay between the exchange field of the ferromagnet at the superconductor–ferromagnet interface and the presence of a non-homogeneous magnetic field. There exists a large number of proposed mechanisms to create this non-homogeneous field, including, but not limited to, domain walls at the interface, spin–orbit interactions in the superconductor, a rotating magnetization in the ferromagnet close to the interface, and local magnetic impurities. Although we can only speculate which of these mechanisms is effective in our case, the large spread in the critical currents observed in our experiments suggests that the process creating the inhomogeneous field—and thus the process responsible for the singlet–triplet conversion—is poorly defined. This suggests that the formation of the interface plays a crucial role.

**Author Information** Reprints and permissions information is available at npg.nature.com/reprintsandpermissions. The authors declare no competing financial interests.

Correspondence and requests for materials should be addressed to R.S.K. (r.s.keizer@tnw.tudelft.nl) or T.M.K. (t.m.klapwijk@tnw.tudelft.nl).

We thank I. van Dijk, M. G. Flokstra, H. T. Man and S. Russo for stimulating interactions. This work is part of the research programme of the Stichting voor Fundamenteel Onderzoek der Materie (FOM), which is financially supported by the Nederlandse Organisatie voor Wetenschappelijk Onderzoek (NWO). The work at the University of Alabama was supported by the National Science Foundation as part of an MRSEC grant.

**Figure 1 Basic aspects of the experimental system. a**, Simplified view of the spin dependent density-of-states (DOS) of $CrO_2$. At the Fermi level, there is a gap in the DOS for spin-down, while the spin-up band is metallic, leading to a fully spin-polarized conductor. The absence of spin-down states rules out spin-flip scattering in the transport. $E_F$, Fermi energy. **b**, Schematic illustration of the studied devices. The half-metallic $CrO_2$ (100) single crystal thin film (100 nm thick) is epitaxially grown on top of an (insulating) $TiO_2$ (100) substrate by chemical vapour deposition[23]. Then, patterns are defined in an organic resist mask with conventional electron beam lithography. Before the deposition of the *s*-wave superconductor NbTiN (ref. 5) contacts, the surface not covered by the mask is sputter-cleaned with an Ar plasma to remove the natural oxide $Cr_2O_3$ terminating the surface in order to obtain high quality contacts to the $CrO_2$. Note that this fabrication procedure precludes spurious connections between the NbTiN electrodes, as independently confirmed by scanning electron microscopy. **c**, Scanning electron micrograph of a typical final device. **d**, Illustration of the alignment of the current direction with respect to the magnetization axes. $\psi$ is the angle of the applied magnetic field **H** with respect to the current direction **I**, and $\theta$ is the direction of the magnetization **M**.

**Figure 2 Observed superconducting transport properties of the superconductor–$CrO_2$–superconductor system. a**, Typical current–voltage (*I–V*) characteristic at temperature $T$=1.6 K: a zero resistance supercurrent branch is clearly visible (for larger critical currents the current–voltage characteristic is hysteretic, see inset). Similar data have been observed in 10 different samples, some of which had several devices in series. From device to device a spread of critical current of less than 2 orders of magnitude is found. The magnitude of $I_cR_N$, (the product of the critical current, $I_c$, and the normal state resistance, $R_N$) is for all junctions smaller than 4 mV (twice the estimated gap size of the NbTiN), and typically 10–300 µV, for nominal junction lengths, $L$, of 0.3-1 µm. **b**, Critical current as a function of temperature for three devices. **c**, Critical current as a function of external magnetic field, applied in the plane of the film, for the device used in **a**. The angle $\psi$ between the direction of the current and the field is 90°. The raw data are shown in the inset with a trace for increasing (up-sweep) and one for decreasing (down-sweep) magnetic field strength, demonstrating that we observe hysteresis. In the main figure the measurements for decreasing field strength are shifted by $\mu_0 H = 45$ mT (with the permeability of vacuum, $\mu_0$) with respect to those for increasing field to correct for the hysteresis. The main figure clearly shows oscillations in the critical current with the applied magnetic flux through the junction. For technical reasons no measurements are performed in perpendicular field.

**Figure 3 Control of the critical current by changing the magnetization orientation.** The hysteresis loop in the critical current for a sweep from low to high fields ('major loop' from V to VI through I and II and back through III and IV) is different from the hysteresis for a sweep from low to moderately high fields ('minor loop' from V to II and back), with the Roman capitals indicating the axes of the biaxial symmetry (inset). These data illustrate that the $CrO_2$ behaves as a single domain. The data are for clarity presented as a 5-point average.

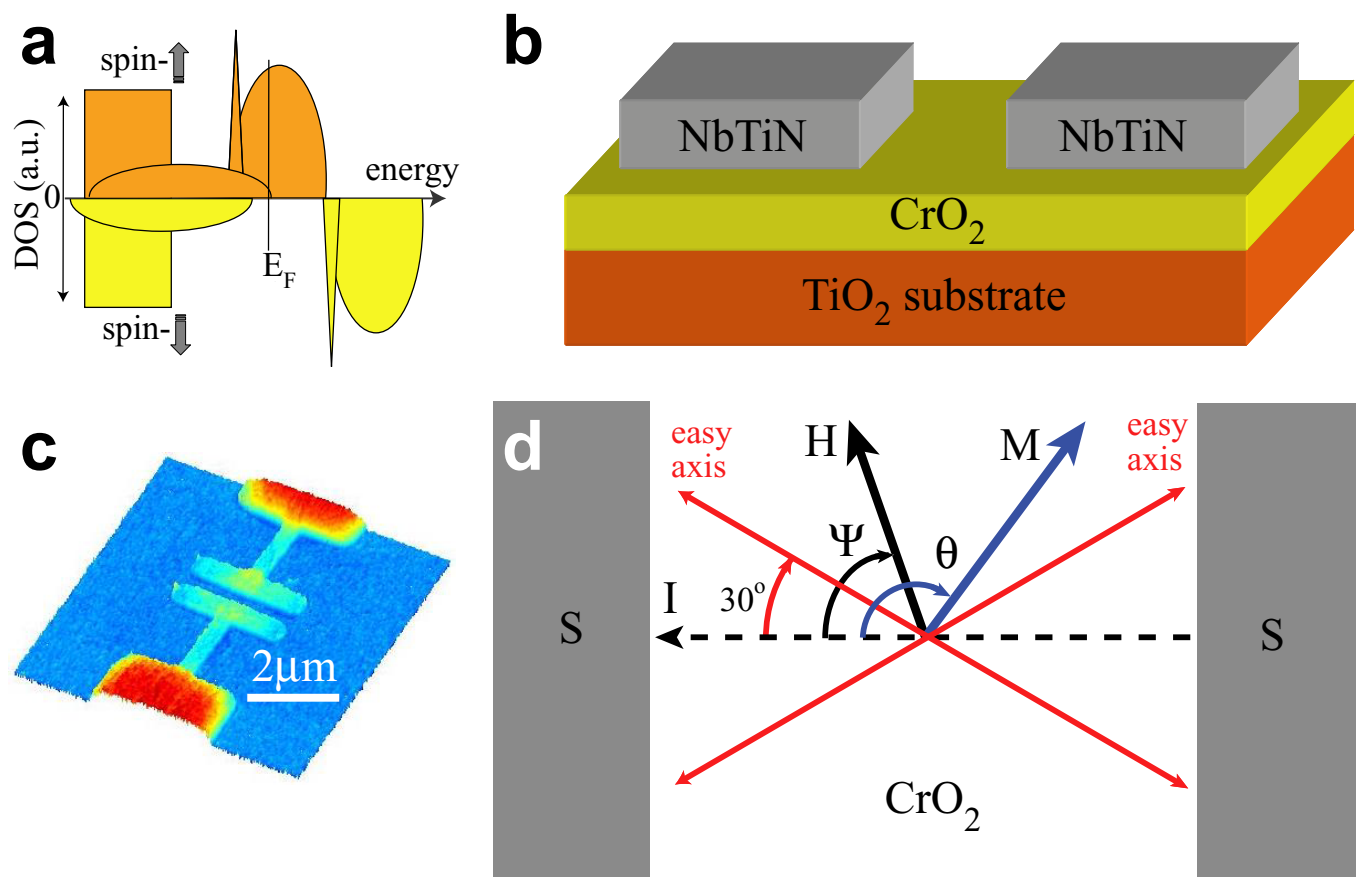

Figure 1 Keizer *et al.*

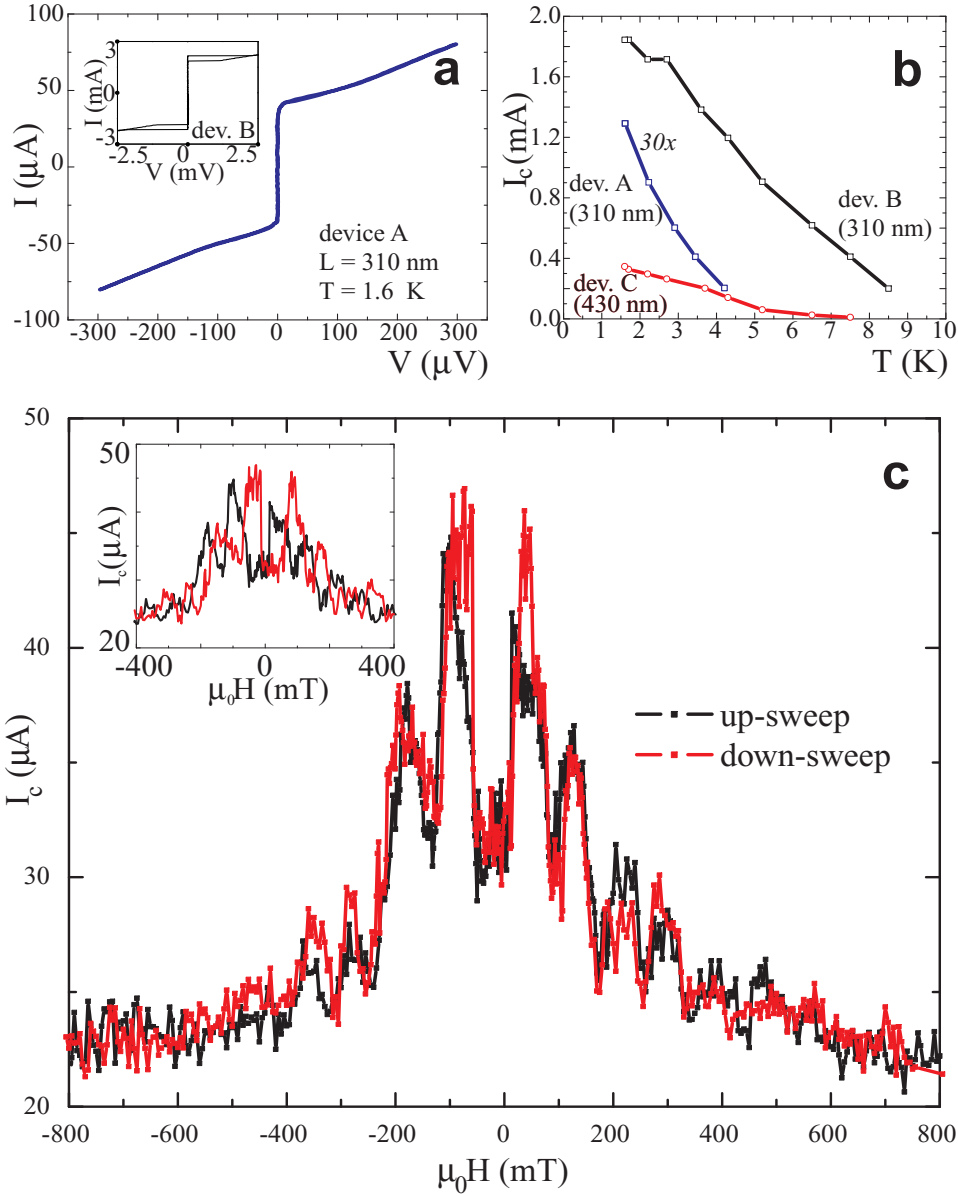

Figure 2 Keizer *et al.*

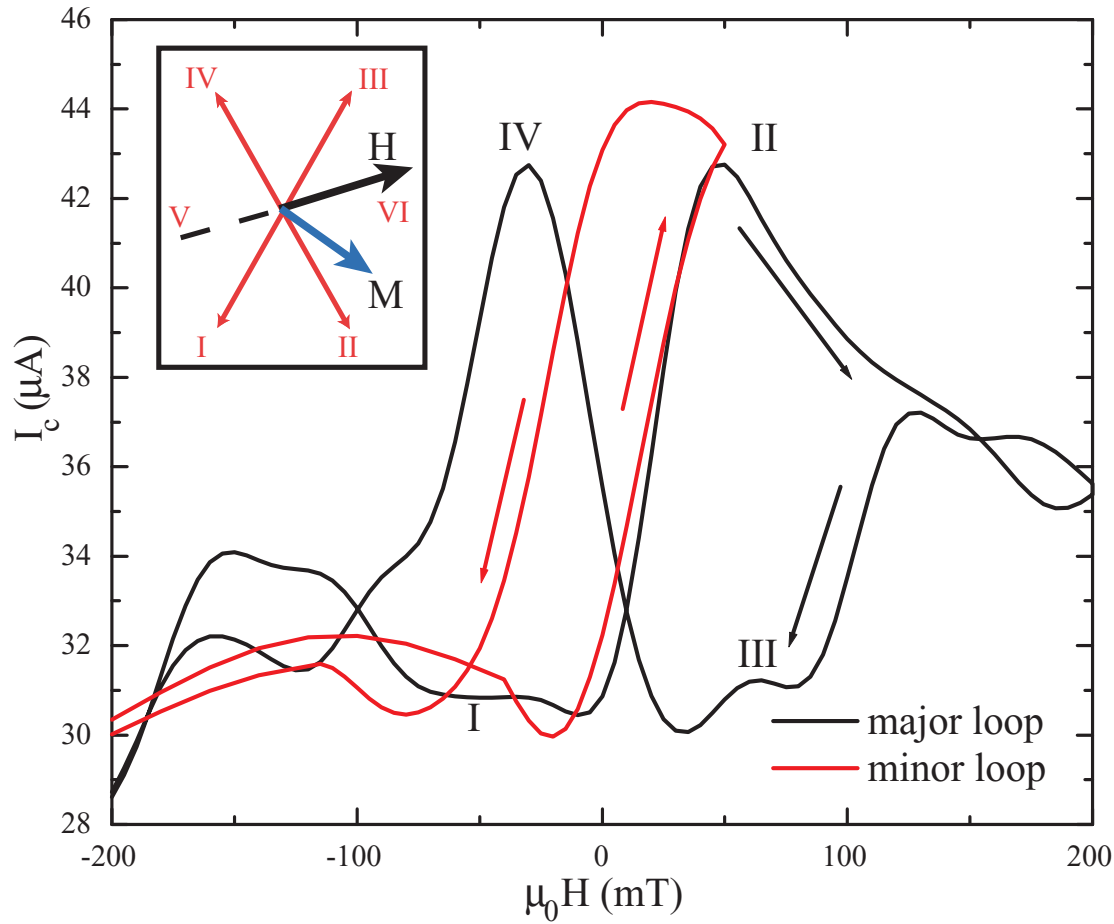

Figure 3 Keizer *et al.*